\documentstyle[multicol,aps,pre,epsf]{revtex}

\begin{document}
\title{Wealth Distributions in Models of Capital Exchange}

\author{S.~Ispolatov, P.~L.~Krapivsky, and S.~Redner}
\address{Center for Polymer Studies and Department of Physics,
Boston University, Boston, MA 02215}

\maketitle
\begin{abstract}

A dynamical model of capital exchange is introduced in which a specified
amount of capital is exchanged between two individuals when they meet.
The resulting time dependent wealth distributions are determined for a
variety of exchange rules.  For ``greedy'' exchange, an interaction
between a rich and a poor individual results in the rich taking a
specified amount of capital from the poor.  When this amount is
independent of the capitals of the two traders, a mean-field analysis
yields a Fermi-like scaled wealth distribution in the long-time limit.
This same distribution also arises in greedier exchange processes, where
the interaction rate is an increasing function of the capital difference
of the two traders.  The wealth distribution in multiplicative
processes, where the amount of capital exchanged is a finite fraction of
the capital of one of the traders, are also discussed.  For random
multiplicative exchange, a steady state wealth distribution is reached,
while in greedy multiplicative exchange a non-steady power law wealth
distribution arises, in which the support of the distribution
continuously increases.  Finally, extensions of our results to arbitrary
spatial dimension and to growth processes, where capital is created in
an interaction, are presented.  

\vskip 0.1in 


\end{abstract}
\begin{multicols}{2}

\section{INTRODUCTION}

Economics underlies many of our day-to-day activities, and yet,
remarkably, resists axiomatization and first-principles explanations.
However, the recent applications of ideas developed in statistical
physics, such as scaling, self-organization, landscape paradigms, {\it
etc.} may help establish a conceptual framework for the scientific
analysis of economic activities\cite{mandel,pwa,kauf,bak}.  In this
spirit of simplicity and concreteness, we introduce capital exchange
models as an attempt to account for the wealth distribution of a
population.  While there is considerable data available on this
phenomenon\cite{dist}, there does not appear to be a simple and causal
explanation of the observed distributions.  The basis of our modeling
is that the elemental kernel of economic activity in a dynamic economy
is the interaction between two individuals which results in the
rearrangement of their capital.  Through repeated two-body interactions
between pairs of traders, a global wealth distribution develops
and we wish to understand how generic features of this distribution
depend on the nature of the two-body interaction.

Our models have the obvious shortcomings of being oversimplified and on
focusing on only one mechanism among the myriad of factors that
influence individual wealth.  Based on everyday experience, however, it
may be argued that at an elemental level much economic activity is
essentially the trading of capital between two individuals.  The
potential utility of our models is that they yield realistic wealth
distributions for certain exchange rules, as well suggesting avenues for
potentially fruitful developments.

In the next section, we first treat ``additive'' processes in which a
fixed amount of capital is exchanged in an interaction between two
individuals, independent of their initial capital\cite{melzak}.  While
the restriction of fixed capital is not realistic, the resulting models
are soluble and provide a natural starting point for
economically-motivated generalizations.  Within a (mean-field) rate
equation description, we find that the wealth distribution generally
exhibits scaling, from which both the time dependence of the average
wealth and the shape of the scaled wealth distribution can be obtained.
As a preliminary, we give an exact solution for random exchange, where
either trader is equally likely to profit in an interaction.  The
scaling approach is then applied to ``greedy'' exchange, in which the
richer person always wins in an interaction.  For this system, the
resulting wealth distribution closely resembles the occupancy
distribution of an ideal Fermi gas.

We next consider ``multiplicative'' processes in Sec.\ III, in which the
amount of capital exchanged is a fixed fraction of the current capital
of one of the traders.  The motivation for this model is the observation
that fractional exchange typically underlies many economic transactions
-- for example, a loan at the end of its repayment period.  For a broad
range of model parameters, realistic forms of the wealth distribution
result.  In the interesting case of greedy multiplicative exchange, the
distribution is a power with a cutoff at small wealth that decays
exponentially in time and a large wealth cutoff that grows linearly.
In Sec.\ IV, we then discuss the behavior of our wealth exchange models
in arbitrary spatial dimension, when diffusion is the transport
mechanism which brings trading partners together.  As might be
anticipated, when the spatial dimension $d<2$, dimension-dependent
behavior is found for the time evolution of the wealth distribution.
Finally, in Sec.\ V, we summarize and also briefly mention a toy model
for economic growth in which capital is produced in an elemental
two-body interaction.  The relation between capital production and
capital conserving models are also outlined.

\section{ADDITIVE CAPITAL EXCHANGE}

Consider a population of traders, each of which possesses a certain
amount of capital which is assumed to be quantized in units of minimal
capital.  Taking this latter quantity as the basic unit, the fortune of
an individual is restricted to the integers.  The wealth of the
population evolves by the repeated interaction of random pairs of
traders.  In each interaction, one unit of capital is transferred
between the trading partners.  To complete the description, we specify
that if a poorest individual (with one unit of capital) loses all
remaining capital by virtue of a ``loss'', the bankrupt individual is
considered to be economically dead and no longer participates in
economic activity.

In the following subsections, we consider three specific realizations of
additive capital exchange.  In ``random'' exchange, the direction of the
capital exchange is independent of the relative capital of the traders.
While this rule has little economic basis, the resulting model is
completely soluble and thus provides a helpful pedagogical starting
point.  We next consider ``greedy'' exchange in which a richer person
takes one unit of capital from a poorer person in a trade.  Such a rule
is a reasonable starting point for describing exploitive economic
activity.  Finally, we consider a more heartless version -- ``very
greedy'' exchange -- in which the rate of exchange is proportional to
the difference in capital between the two interacting individuals.
These latter two cases can be solved by a scaling approach.  The primary
result is that the scaled wealth distribution resembles a
finite-temperature Fermi distribution, with an effective temperature
that goes to zero in the long-time limit.

\subsection{Random Exchange}

In this process, one unit of capital is exchanged between trading
partners, as represented by the reaction scheme $(j,k)\to (j\pm 1,k\mp
1)$.  Let $c_k(t)$ be the density of individuals with capital $k$.
Within a mean-field description, $c_k(t)$ evolves according to

\begin{equation}
\label{ckran}
{dc_k(t)\over dt}=N(t)\left[c_{k+1}(t)+c_{k-1}(t)-2c_k(t)\right],
\end{equation}
with $N(t)\equiv M_0(t)=\sum_{k=1}^\infty c_k(t)$ the population
density.  The first two terms account for the gain in $c_k(t)$ due to
the interactions $(j,k+1)\to (j+1,k)$ and $(j,k-1)\to (j-1,k)$,
respectively, while the last term accounts for the loss in $c_k(t)$ due
to the interactions $(j,k)\to (j\pm 1,k\mp 1)$

By defining a modified time variable,

\begin{equation}
\label{T}
T=\int_0^t dt' N(t'),
\end{equation}
Eq.~(\ref{ckran}) is reduced to the discrete diffusion equation

\begin{equation}
\label{diff}
{dc_k(T)\over dT}=c_{k+1}(T)+c_{k-1}(T)-2c_k(T).
\end{equation}
The rate equation for the poorest density has the slightly different
form, $dc_1/dT=c_2-2c_1$, but may be written in the same form as
Eq.~(\ref{diff}) if we impose the boundary condition $c_0(T)=0$.

Eq.~(\ref{diff}) may be readily solved for arbitrary initial
conditions\cite{feller}.  For illustrative purposes, let us assume that
initially all individuals have one unit of capital,
$c_k(0)=\delta_{k1}$.  The solution to Eq.~(\ref{diff}) subject to these
initial and boundary conditions is

\begin{equation}
\label{cksol}
c_k(T)=e^{-2T}\left[I_{k-1}(2T)-I_{k+1}(2T)\right],
\end{equation}
where $I_n$ denotes the modified Bessel function of order $n$
\cite{bender}.  Consequently, the total density $N(T)$ is
\begin{equation}
\label{Nsol}
N(T)=e^{-2T}\left[I_0(2T)+I_1(2T)\right].
\end{equation}

To re-express this exact solution in terms of the physical time $t$, we
first invert Eq.~(\ref{T}) to obtain $t(T)=\int_0^T dT'/N(T')$, and then
eliminate $T$ in favor of $t$ in the solution for $c_k(T)$.  For
simplicity and concreteness, let us consider the long-time limit.  From
Eq.~(\ref{cksol}),

\begin{equation}
\label{ckT}
c_k(T)\simeq {k\over \sqrt{4\pi T^3}}\,
\exp\left(-{k^2\over 4T}\right),
\end{equation}
and from Eq.~(\ref{Nsol}),

\begin{equation}
\label{NT}
N(T)\simeq (\pi T)^{-1/2}. 
\end{equation}
Eq.~(\ref{NT}) also implies $t\simeq {2\over 3}\sqrt{\pi T^3}$ 
which gives

\begin{equation}
\label{Nasym}
N(t)\simeq \left({2\over 3\pi t}\right)^{1/3},
\end{equation}
and

\begin{equation}
\label{ckasym}
c_k(t)\simeq {k\over 3t}\,
\exp\left[-\left({\pi\over 144}\right)^{1/3}\,{k^2\over t^{2/3}}\right].
\end{equation}

Note that this latter expression may be written in the scaling form
$c_k(t)\propto N^2x\,e^{-x^2}$, with the scaling variable $x\propto kN$.
One can also confirm that the scaling solution represents the basin of
attraction for almost all exact solutions.  Indeed, for any initial
condition with $c_k(0)$ decaying faster than $k^{-2}$, the system
reaches the scaling limit $c_k(t)\propto N^2x\,e^{-x^2}$.  On the other
hand, if $c_k(0)\sim k^{-1-\alpha}$, with $0<\alpha<1$, such an initial
state converges to an alternative scaling limit which depends on
$\alpha$, as discussed, {\it e.\ g.}, in Ref.~\cite{dgy}.   These
solutions exhibit a slower decay of the total density, $N\sim
t^{-\alpha/(1+\alpha)}$, while the scaling form of the wealth
distribution is

\begin{equation}
\label{cklong}
c_k(t)\sim N^{2/\alpha}{\cal C}_{\alpha}(x), \quad 
x\propto kN^{1/\alpha},
\end{equation}
with the scaling function 

\begin{equation}
\label{alpha}
{\cal C}_{\alpha}(x)=e^{-x^2}\int_0^\infty 
du\,{e^{-u^2}\sinh(2ux)\over u^{1+\alpha}}.
\end{equation}
Evaluating the integral by the Laplace method gives an asymptotic
distribution which exhibits the same $x^{-1-\alpha}$ as the initial
distribution.  This anomalous scaling in the solution to the diffusion
equation is a direct consequence of the extended initial condition.
This latter case is not physically relevant, however, since the extended
initial distribution leads to a divergent initial wealth density.

\subsection{Greedy Exchange}

In greedy exchange, when two individuals meet, the richer person takes
one unit of capital from the poorer person, as represented by the
reaction scheme $(j,k)\to (j+1,k-1)$ for $j\geq k$.  In the rate
equation approximation, the densities $c_k(t)$ now evolve according to

\begin{equation}
\label{ck}
{dc_k\over dt}=c_{k-1}\sum_{j=1}^{k-1}c_j+c_{k+1}\sum_{j=k+1}^\infty c_j
-c_k N-c_k^2.
\end{equation}
The first two terms account for the gain in $c_k(t)$ due to the
interaction between pairs of individuals of capitals $(j,k-1)$, with
$j<k$ and $(j,k+1)$ with $j>k$, respectively.  The last two terms
correspondingly account for the loss of $c_k(t)$.  One can check that
the wealth density $M_1\equiv\sum_{k=1}^\infty k\,c_k(t)$ is conserved and
that the population density obeys

\begin{equation}
\label{Nt}
{dN\over dt}=-c_1 N.
\end{equation}

Eqs.~(\ref{ck}) are conceptually similar to the Smoluchowski equations
for aggregation with a constant reaction rate\cite{sm}.  Mathematically,
however, they appear to be more complex and we have been unable to
solve them analytically.  Fortunately, Eq.~(\ref{ck}) is amenable to a
scaling solution\cite{ernst}.  For this purpose, we first re-write
Eq.~(\ref{ck}) as

\begin{eqnarray}
\label{ckt}
{dc_k\over dt}=&-&c_k(c_k+c_{k+1})+N(c_{k-1}-c_k)\nonumber\\
               &+&(c_{k+1}-c_{k-1})\sum_{j=k}^\infty c_j.
\end{eqnarray}
Taking the continuum limit and substituting the scaling ansatz

\begin{equation}
c_k(t)\simeq N^2{\cal C}(x), \quad{\rm with}\qquad x=kN,
\label{scal}
\end{equation}
transforms Eqs.~(\ref{Nt}) and (\ref{ckt}) to

\begin{equation}
\label{Nscal}
{dN\over dt}=-{\cal C}(0) N^3,
\end{equation}
and 

\begin{equation}
\label{cx}
{\cal C}(0)[2{\cal C}+x{\cal C}']=2{\cal C}^2
+{\cal C}'\left[1-2\int_x^\infty dy{\cal C}(y)\right],
\end{equation}
where ${\cal C}'=d{\cal C}/dx$.  Note also that the scaling function
must obey the integral relations

\begin{equation}
\label{rel}
\int_0^\infty dx\,{\cal C}(x)=1, \quad{\rm and}\qquad
\int_0^\infty dx\,x\,{\cal C}(x)=1.
\end{equation}
The former follows from the definition of the density, $N=\sum
c_k(t)\simeq N\int dx\,{\cal C}(x)$, while the latter follows if we set,
without loss of generality, the (conserved) wealth density equal to
unity, $\sum_k kc_k(t)=1$.

Introducing ${\cal B}(x)=\int_0^x dy\,{\cal C}(y)$ recasts
Eq.~(\ref{cx}) into ${\cal C}(0)[2{\cal B}'+x{\cal B}'']=2{\cal B}'^2
+{\cal B}''[2{\cal B}-1]$.  Integrating twice gives $[{\cal C}(0)x-{\cal
B}][{\cal B}-1]=0$, with solution ${\cal B}(x)={\cal C}(0)x$ for
$x<x_{\rm f}$ and ${\cal B}(x)=1$ for $x\geq x_{\rm f}$, from which we
conclude that the scaled wealth distribution ${\cal C}(x)={\cal B}'(x)$
coincides with the zero-temperature Fermi distribution;

\begin{equation}
\label{C}
{\cal C}(x)=\cases{{\cal C}(0),    &$x<x_{\rm f}$;\cr
                   0,              &$x\geq x_{\rm f}$.\cr}
\end{equation}
Hence the scaled profile has a sharp front at $x=x_{\rm f}$, with
$x_{\rm f}=1/{\cal C}(0)$ found by matching the two branches of the
solution for ${\cal B}(x)$.  Making use of the second integral relation
(\ref{rel}) gives ${\cal C}(0)=1/2$ and thereby closes the solution.
Thus the unscaled wealth distribution $c_k(t)$ reads

\begin{equation}
\label{fermi}
c_k(t)=\cases{1/(2t),       &$k<2\sqrt{t}$;\cr
              0,            &$k\geq 2\sqrt{t}$;\cr}
\end{equation}
and the total density is $N(t)=t^{-1/2}$.  

\begin{figure}
\narrowtext
\epsfxsize=3in
\epsfbox{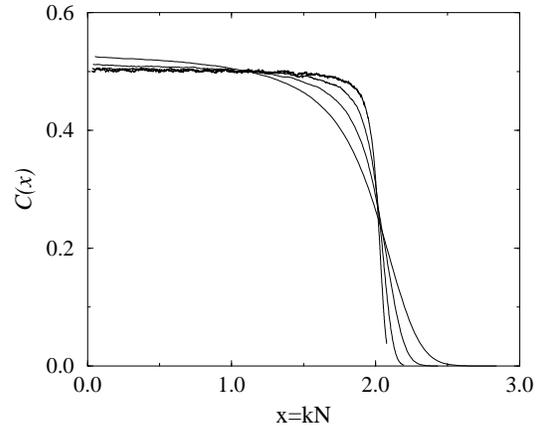}
\vskip 0.2in
\caption{Simulation results for the wealth distribution in greedy
additive exchange based on 2500 configurations for $10^6$ traders.
Shown are the scaled distributions ${\cal C}(x)$ versus $x=kN$ for
$t=1.5^n$, with $n=18$, 24, 30, and 36; these steepen with increasing
time.  Each data set has been averaged over a range of $\approx 3\%$ of
the data points to reduce fluctuations.
\label{fig1}}
\end{figure}

We checked these predictions by numerical simulations (Fig.~1).  In the
simulation, two individuals are randomly chosen to undergo greedy
exchange and this process is repeated.  When an individual reaches zero
capital he is eliminated from the system, and the number of active
traders is reduced by one.  After each reaction, the time is incremented
by the inverse of the number of active traders.  While the mean-field
predictions are substantially corroborated, the scaled wealth
distribution for finite time actually resembles a finite-temperature
Fermi distribution (Fig.~1).  As time increases, the wealth distribution
becomes sharper and approaches Eq.~(\ref{fermi}).  In analogy with the
Fermi distribution, the relative width of the front may be viewed as an
effective temperature.  Thus the wealth distribution is characterized by
two scales; one of order $\sqrt{t}$ characterizes the typical wealth of
active traders and a second, smaller scale which characterizes the width
of the front\cite{bril}.

To quantify the spreading of the front, let us include the next
corrections in the continuum limit of the rate equations,
Eq.~(\ref{ckt}).  This gives,

\begin{equation}
\label{cont}
{\partial c\over \partial t}=
2{\partial \over \partial k}\left[c\int_k^\infty dj\,c(j)\right]
- c{\partial c\over \partial k}
-N{\partial c\over \partial k}
+{N\over 2}{\partial^2 c\over \partial k^2}.
\end{equation}
Here the second and fourth terms on the right-hand side represent the
next corrections.  Since the ``convective'' (third) term determines the
location of the front to be at $k_{\rm f}=2\sqrt{t}$, it is natural to
expect that the (diffusive) fourth term describes the spreading of the
front.  The term $c{\partial c\over\partial k}$ turns out to be
negligible in comparison to the diffusive spreading term and is
henceforth neglected. 

The dominant convective term can be removed by transforming to a frame
of reference which moves with the front, namely, $k\to K=k-2\sqrt{t}$.
Among the remaining terms in the transformed rate equation, the width of
the front region $W$ can now be determined by demanding that the
diffusion term has the same order of magnitude as the reactive terms,
that is, $N{\partial^2 c\over \partial k^2}\sim c^2$.  This implies 
$W\sim\sqrt{N/c}$.  Combining this with $N=t^{-1/2}$ and $c\sim t^{-1}$
gives $W\sim t^{1/4}$, or a relative width $w=W/k_{\rm f}\sim t^{-1/4}$.
This suggests the appropriate scaling ansatz for the front region is

\begin{equation}
\label{scalxi}
c_k(t)={1\over t}X(\xi), \quad
\xi={k-2\sqrt{t}\over t^{1/4}}.
\end{equation}
Substituting this ansatz into Eq.~(\ref{cont}) gives a non-linear single
variable integro-differential equation for the scaling function
$X(\xi)$.  Together with the appropriate boundary conditions, this
represents, in principle, a more complete solution to the wealth
distribution.  However, the essential scaling behavior of the
finite-time spreading of the front is already described by
Eq.~(\ref{scalxi}), so that solving for $X(\xi)$ itself does not provide
additional scaling information.  Analysis of our data by several
rudimentary approaches gives $w\sim t^{-\alpha}$ with $\alpha\approx
1/5$.  We attribute this discrepancy to the fact that $w$ is obtained by
differentiating ${\cal C}(x)$, an operation which generally leads to an
increase in numerical errors.

\subsection{Very Greedy Exchange}

We now consider the variation in which events occur at a rate equal to
the difference in capital of the two traders.  That is, an individual is
more likely to take capital from a much poorer person rather than from
someone of slightly less wealth.  For this ``very greedy'' exchange, the
corresponding rate equations are

\begin{eqnarray}
\label{cknew}
{dc_k\over dt}&=&c_{k-1}\sum_{j=1}^{k-1}(k-1-j)c_j
+c_{k+1}\sum_{j=k+1}^\infty (j-k-1)c_j\nonumber\\
&-&c_k\sum_{j=1}^{\infty}|k-j|c_j,
\end{eqnarray}
while the total density obeys

\begin{equation}
\label{Nnew}
{dN\over dt}=-c_1(1-N),
\end{equation}
under the assumption that the (conserved) total wealth density is set
equal to one, $\sum kc_k=1$.

These rate equations can be solved by again applying scaling.  For this
purpose, it is first expedient to rewrite the rate equations as

\begin{eqnarray}
\label{ckaggr}
{dc_k\over dt}&=&(c_{k-1}-c_k)\sum_{j=1}^{k-1}(k-j)c_j
-c_{k-1}\sum_{j=1}^{k-1}c_j\\
&+&(c_{k+1}-c_k)\sum_{j=k+1}^\infty (j-k)c_j
-c_{k+1}\sum_{j=k+1}^{\infty}c_j\nonumber.
\end{eqnarray}
Taking the continuum limit gives

\begin{equation}
\label{ckcont}
{\partial c\over \partial t}={\partial c\over \partial k}
-N{\partial \over \partial k}(kc).
\end{equation}
We now substitute the scaling ansatz, Eq.~(\ref{scal}), to yield

\begin{equation}
\label{cxaggr}
{\cal C}(0)[2{\cal C}+x{\cal C}']=(x-1){\cal C}'+{\cal C},
\end{equation}
and 
\begin{equation}
\label{Naggr}
{dN\over dt}=-{\cal C}(0) N^2.
\end{equation}

Solving the above equations gives $N\simeq [{\cal C}(0)t]^{-1}$ and

\begin{equation}
\label{Caggr}
{\cal C}(x)=(1+\mu)(1+\mu x)^{-2-1/\mu}, 
\end{equation}
with $\mu={\cal C}(0)-1$.  It may readily be verified that this
expression for ${\cal C}(x)$ satisfies both integral relations of
Eq.~(\ref{rel}).  The scaling approach has thus found a family of
solutions which are parameterized by $\mu$, and additional information
is needed to determine which of these solutions is appropriate for our
system.  For this purpose, note that Eq.~(\ref{Caggr}) exhibits
different behaviors depending on the sign of $\mu$.  When $\mu>0$, there
is an extended non-universal power-law distribution, while for $\mu=0$
the solution is the pure exponential, ${\cal C}(x)=e^{-x}$.  These
solutions may be rejected because the wealth distribution cannot extend
over an unbounded domain if the initial wealth extends over a finite
range.

The accessible solutions therefore correspond to $-1<\mu<0$, where the
distribution is compact and finite, with ${\cal C}(x)\equiv 0$ for
$x\geq x_{\rm f}=-\mu^{-1}$.  To determine the true solution, let us
re-examine the continuum form of the rate equation, Eq.~(\ref{ckcont}).
From naive power counting, the first two terms are asymptotically
dominant and they give a propagating front with $k_{\rm f}$ exactly
equal to $t$.  Consequently, the scaled location of the front is given
by $x_{\rm f}=Nk_{\rm f}$.  Now the result $N\simeq [{\cal C}(0)t]^{-1}$
gives $x_{\rm f}=1/{\cal C}(0)$.  Comparing this expression with the
corresponding value from the scaling approach, $x_{\rm f}=[1-{\cal
C}(0)]^{-1}$, selects the value ${\cal C}(0)=1/2$.  Remarkably, this
scaling solution coincides with the Fermi distribution that found for
the case of constant interaction rate.  Finally, in terms of the
unscaled variables $k$ and $t$, the wealth distribution is

\begin{equation}
\label{fermi2}
c_k(t)=\cases{4/t^2,    &$k<t$;\cr
              0,        &$k\geq t$.\cr}
\end{equation}
Following the same reasoning as the previous section, this discontinuity
is smoothed out by diffusive spreading.

Another interesting feature is that if the interaction rate is
sufficiently greedy, ``gelation'' occurs \cite{gel}, whereby a finite
fraction of the total capital is possessed by a single individual.  For
interaction rates, or kernels $K(j,k)$ between individuals of capital
$j$ and $k$ which do not give rise to gelation, the total density
typically varies as a power law in time, while for gelling kernels
$N(t)$ goes to zero at some finite time.  At the border between these
regimes $N(t)$ typically decays exponentially in time\cite{ernst,gel}.
We seek a similar transition in behavior for the capital exchange model
by considering the rate equation for the density

\begin{equation}
\label{Ngen}
{dN\over dt}=-c_1\sum_{k=1}^\infty K(1,k)c_k.
\end{equation}
For the family of kernels with $K(1,k)\sim k^\nu$ as $k\to\infty$,
substitution of the scaling ansatz gives $\dot N\sim -N^{3-\nu}$.  Thus
$N(t)$ exhibits a power-law behavior $N\sim t^{1/(2-\nu)}$ for $\nu<2$
and an exponential behavior for $\nu=2$.  Thus gelation should arise for
$\nu>2$.

\section{MULTIPLICATIVE CAPITAL EXCHANGE}

We have thus far focused on additive processes in which the amount of
capital transferred in a two-body interaction is fixed.  This leads to
the unrealistic feature of a vanishing density of active traders in the
long time limit, as an individual who possesses the minimal amount of
capital loses all assets in an unfavorable interaction.  In many
economic transactions, however, the amount of capital transferred is a
finite fraction of the initial capital of one of the participants.  This
observation motivates us to consider capital exchange models with
exactly this multiplicative property.  A simple realization which
preserves both the number of participants and the total capital is the
reaction scheme $(x,y)\to (x-\alpha x, y+\alpha x)$.  Here $0<\alpha<1$
represents the fraction of loser's capital which is gained by the
winner.  In this process, the capital of any individual remains
non-zero, although it can become vanishingly small.

In the following, we consider the cases of random exchange, where the
winner may equally likely be the richer or the poorer of the two
traders, and greedy exchange, where only the richer of the two traders
profits in the interaction.  The former system quickly reaches a steady
state, while the latter gives rise to a non-stationary power-law
distribution of wealth.

\subsection{Random Exchange}

To determine the rate equation for random multiplicative exchange, it is
expedient to first write an integral form of the equation, for which the
origin of the various terms is clear.  This rate equation is
\begin{eqnarray}
\label{consint}
{\partial c(x)\over \partial t}={1\over 2}
&&\int\int dy\,dz\,c(y)c(z)\times [-\delta(x-z)-\delta(x-y)\nonumber\\
&&+\delta(y(1-\alpha)-x)+\delta(z+\alpha y-x)].
\end{eqnarray}
The first two terms account for the loss of $c(x)$ due to the
interaction of an individual of capital $x$.  The next term accounts for
the gain in $c(x)$ by the losing interaction
$(x/(1-\alpha),y)\to(x,y+\alpha x/(1-\alpha))$.  The last term also
accounts for gain in $c(x)$ by the profitable interaction $(y,x-\alpha
y)\to(y(1-\alpha),x)$.  By integrating over the delta functions, this
rate equation reduces to

\begin{eqnarray}
\label{cons}
{\partial c(x)\over \partial t}= -c(x)&+&{1\over 2(1-\alpha)}\,
c\left({x\over 1-\alpha}\right)\nonumber\\
&+&{1\over 2\alpha}\int_0^x dy\,c(y)\,c\left({x-y\over \alpha}\right),
\end{eqnarray}
where the total density is set equal to one.  In this form, the rate
equation describes a diffusive-like process on a logarithmic scale,
except that the (third) term, which describes hopping to the right, is
non-local and two-body in character.  

To help understand the nature of the resulting wealth distribution, let
us first consider the moments, $M_n(t)\equiv\int_0^\infty dx\,x^n
c(x,t)$.  From Eq.~(\ref{cons}) one can straightforwardly verify that
the first two moments, $M_0$ and $M_1$, the population and wealth
densities, respectively, are conserved.  Without loss of generality, we
choose $M_0=1$ and $M_1=M$.  The equation of motion for the second
moment is

\begin{equation}
\label{mom2}
{d M_2(t)\over dt}=-\alpha(1-\alpha)M_2(t)+\alpha M^2
\nonumber
\end{equation}
with solution

\begin{equation}
\label{m2}
M_2(t)={M^2\over 1-\alpha}
+\left[M_2(0)-{M^2\over 1-\alpha}\right]\,e^{-\alpha(1-\alpha)t}.
\end{equation}
Similarly, higher moments also exhibit exponential convergence to
constant values, so that the wealth distribution approaches a steady
state.  The mechanism for this steady state is simply that the typical
size of a profitable interaction is likely to be much smaller than an
unprofitable interaction for a rich individual, while the opposite holds
for a poor individual.  This bias prevents the unlimited spread of the
wealth distribution and stabilizes a steady state.

To determine the steady state wealth distribution, we substitute simple
``test'' solutions into the rate equations. By this approach, we find
that the exponential form $c(x)=Be^{-bx}$ satisfies the steady-state
version of the rate equation, Eq.~(\ref{cons}), {\em iff}
$\alpha={1\over 2}$ and $B=b$.  Thus when the winner receives one half
the capital of the loser, the exact steady wealth distribution is a
simple exponential $c(x)=M^{-1}\exp(-x/M)$.  For general $0<\alpha<1$,
the large-$\alpha$ tail is again an exponential, $c(x)\simeq
2b(1-\alpha)e^{-bx}$.  However, for $x\ll 1$, we find, by substitution
and applying dominant balance, that $c(x)\sim x^\lambda$ is the
asymptotic solution, with exponent $\lambda= -1-\ln 2/\ln(1-\alpha)$.

\begin{figure}
\narrowtext
\epsfxsize=3in
\epsfbox{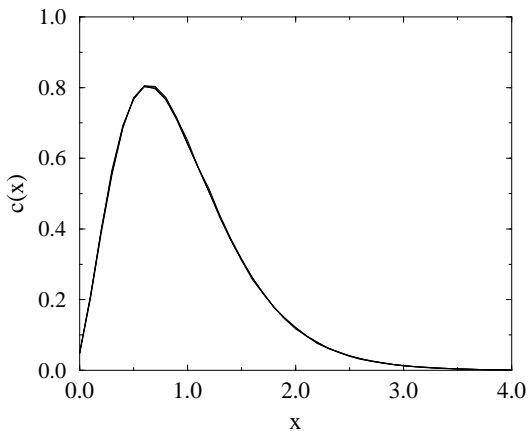}
\vskip 0.2in
\caption{Representative results for the wealth distribution in random 
multiplicative exchange for the case $\alpha=0.25$ based on simulation
of 10 configurations of $10^5$ traders.  Shown are the steady-state
wealth distributions $c(x)$ as a function of wealth $x$ for $t=1.5^n$,
with $n=6$, 8, 10, and 12.  The various curves are indistinguishable.
The predicted $x^{1.409\ldots}$ small-$x$ tail is not resolvable because
of the coarseness of the data binning.
\label{fig2}}
\end{figure}

Interestingly, $\lambda$ is positive when $\alpha<1/2$, so that the
density of the poor is vanishingly small.  A heuristic justification for
this phenomenon is that for $\alpha<1/2$ an unfavorable interaction
leads to a relatively small capital loss, and this loss is more than
compensated for by favorable interactions so that a poor individual has
the possibility of climbing out of poverty.  In the opposite case of
$\alpha>1/2$, unprofitable interactions are sufficiently devastating
that a large and persistent underclass is formed, with a power-law
divergence in the number of poor in the limit of vanishing wealth.

Our simulations substantially confirm these results (Fig.~2).  The scope
of the simulations is less than that in additive processes, since the
number of traders remains fixed, so that CPU time scales linearly in the
simulation time.  In contrast, for additive exchange, the CPU time scale
as $\int^t dt'\, N(t')$, which can be much smaller than t.  Numerically,
we find that the moments $M_n(t)$ quickly converge to equilibrium
values.  The resulting wealth distribution is clearly a simple
exponential for $\alpha=1/2$ and exhibits either a power-law divergence
or a power-law zero as $x\to 0$ for $\alpha>1/2$ and $\alpha<1/2$,
respectively, in agreement with our analytical results.

\subsection{Greedy Exchange}

Parallel to our discussion of additive processes, we now investigate
greedy multiplicative exchange, where only the richer trader profits, as
represented by the reaction $(x,y)\to (x-\alpha x, y+\alpha x)$ for
$x<y$.  Following the same reasoning as that used in the previous
subsection, the rate equation for greedy multiplicative exchange is

\begin{eqnarray}
\label{mge}
{\partial c(x)\over \partial t}&=& -c(x) +
{1\over 1-\alpha}\,c\left({x\over 1-\alpha}\right)
N\left({x\over 1-\alpha}\right) \nonumber\\
&&+{1\over \alpha}\int_{x/(1+\alpha)}^x 
dy\, c(y)\,c\left({x-y\over \alpha}\right),
\end{eqnarray}
where $N(x)=\int_x^\infty dz\,c(z)$ is the population density whose
wealth exceeds $x$.

Numerical simulations of this system show that the wealth distribution
evolves {\it ad infinitum} and that the most of the population
eventually becomes impoverished (Fig.~3).  Note that the discreteness of
our linear data binning lumps the poorest into a single bin at the
origin which is not visible on the double logarithmic scale.  The
pervasive impoverishment arises because greedy exchange causes the poor
to become poorer and the rich to become richer, but wealth conservation
implies that there must be many more poor than rich individuals.  In the
long-time limit therefore, a small fraction of the population possesses
most of the wealth.

To understand these features analytically, first consider the extreme
case of $\alpha=1$ which reduces to classical constant kernel
aggregation\cite{sm}, except for the added feature that individuals of
zero wealth are now included in the distribution.  Consequently, the
scaling form of the wealth distribution in the long time limit is

\begin{equation}
\label{smolu}
c(x,t)\simeq t^{-2}e^{-x/t}+(1-t^{-1})\delta(x).
\end{equation}
The first term is just the scaling solution to constant-kernel
aggregation\cite{sm}, so that the delta function term represents the
population with zero wealth.

\begin{figure}
\narrowtext
\epsfxsize=3in
\epsfbox{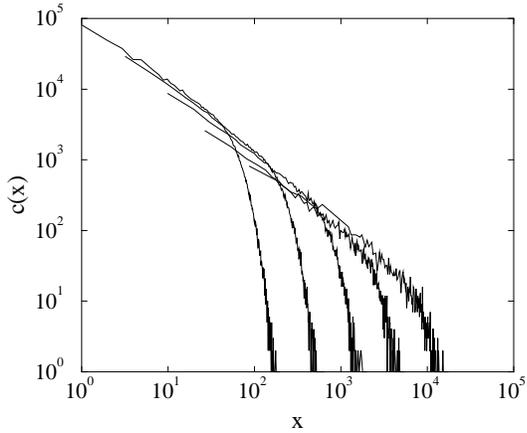}
\vskip 0.2in
\caption{The unnormalized wealth distribution in greedy multiplicative 
exchange for the case $\alpha=0.5$ based on simulation of 10
configurations of $10^5$ traders.  Shown on a double logarithmic scale
are the wealth distributions $c(x)$ as a function of wealth $x$ for
$t=1.5^n$, with $n=7$, 10, 13, and 16.
\label{fig3}}
\end{figure}

For general $0<\alpha<1$ a qualitatively related distribution can be
anticipated which consists of a large impoverished class of negligible
wealth and a much smaller and widely distributed population of wealthy.
To determine this distribution, it is expedient to re-write
Eq.~(\ref{mge}) as

\begin{eqnarray}
\label{mge2}
&&{\partial c(x)\over \partial t}= \int_0^{x/(1+\alpha)}
dz\, c(z)\,[c(x-\alpha z)-c(x)]\\
&&-c(x)N\left({x\over 1+\alpha}\right)+
{1\over 1-\alpha}\,c\left({x\over 1-\alpha}\right)
N\left({x\over 1-\alpha}\right)\nonumber.
\end{eqnarray}
Since Fig.~3 indicates that the wealth distribution is a
power law, we substitute such a form in Eq.~(\ref{mge2}) and find that

\begin{equation}
\label{exact}
c(x,t)={A\over x t}, \quad A=-{1\over \ln(1-\alpha)}
\end{equation}
is a solution.  However, this exact form cannot be realized starting
from any initial condition.  Thus Eq.~(\ref{exact}) should be regarded
as an attractor for the family of solutions to Eq.~(\ref{mge2}) which
begin from a specified initial condition.  Another pathology of
Eq.~(\ref{exact}) is that all moments, $M_n(t)=\int_0^\infty
dx\,x^nc(x,t)$, are divergent.  Note also that the last two terms on the
right-hand side of Eq.~(\ref{mge2}) diverge (although their difference
is regularized to the finite value ${A^2\over xt^2}\,\ln{1+\alpha\over
1-\alpha}$).  These observations suggest that a true solution to
Eq.~(\ref{mge2}) converges to Eq.~(\ref{exact}) only in the scaling
region $x_1(t)<x<x_2(t)$, while a true solution has not yet had time to
extend outside this domain.

To estimate these cutoffs for the scaling region, we evaluate the moments

\begin{eqnarray}
\label{m01}
M_0(t)&\sim & \int_{x_1}^{x_2} dx\,c(x,t)\sim {A\ln (x_2/x_1)\over t}
\nonumber\\
M_1(t)&\sim & \int_{x_1}^{x_2} dx\,xc(x,t)\sim {Ax_2\over t}
\end{eqnarray}
Since $M_0=1$ and $M_1=M$ are constant, one obtains

\begin{equation}
\label{x12}
x_1(t)\sim e^{-t/A}=(1-\alpha)^t, \quad
x_2(t)\propto t.
\end{equation}
The factor $x_1(t)$ clearly gives the wealth of the poorest at time $t$.
Since a losing interaction leads to a reduction of capital by the factor
$(1-\alpha)$, the poorest individual at time $t$ will have capital
$(1-\alpha)^t$ for the monodisperse initial condition,
$c_0(x)=\delta(x-1)$, and a constant reaction rate.  To understand the
upper cutoff, suppose that $x\gg t$.  In this case, the last two terms
on the right-hand side of (\ref{mge2}) are negligible.  In the remaining
term, the expression in the square brackets may be replaced by $\alpha z
{\partial c(x,t)\over \partial x}$, so that the resulting integral is
simply equal to $M$.  With these simplifications, the rate equation
reduces to $c_t+\alpha Mc_x=0$.  This linear wave equation admits the
general solution $c(x,t)=c_0(x-\alpha Mt)$ and suggests the upper cutoff
$x_2(t)\simeq\alpha Mt$, consistent with Eq.~(\ref{x12}).

\begin{figure}
\narrowtext
\epsfxsize=3in
\epsfbox{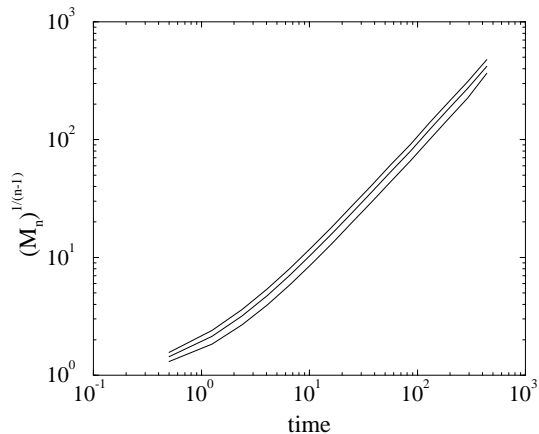}
\vskip 0.2in
\caption{The reduced moments $M_n^{1/(n-1)}$ versus time for greedy
multiplicative exchange for the case $\alpha=0.5$ based on simulation of
10 configurations of $10^5$ traders.  These reduced moments are
predicted to increase linearly in time (see text).
\label{fig4}}
\end{figure}

Note, however, that there is inconsistency in our reasoning, as starting
with the assumption $x\gg t$ leads to an upper cutoff of order $t$.  In
spite of this logical shortcoming, we have verified many of the
resulting quantitative characteristics.  For example, numerical
simulation clearly yields the $1/x$ power-law tail of the wealth
distribution.  Furthermore, if one defines the wealthy as those whose
capital exceeds some threshold $\epsilon$, then Eqs.~(\ref{exact}) and
(\ref{x12}) give the density of wealthy proportional to
$\ln(t/\epsilon)/t$.  It is in this sense that we can view the wealth
distribution as consisting of two components: the wealthy density which
is proportional to $\ln(t/\epsilon)/t$ and a complementary density of
the poor.  Using Eq.~(\ref{exact}) one can also readily determine the
behavior of the moments to be $M_n(t)\sim t^{-1}\int_0^t dx\,x^{n-1}\sim
t^{n-1}$, for $n>1$, or equivalently $M_n^{1/(n-1)}$ should grow
linearly with time, as is observed in our simulations (Fig.~4).  
Least-square fits to the data with the first few data points deleted
clearly indicate a growth rate which is very nearly linear.

\section{ARBITRARY SPATIAL DIMENSION}

We now consider the role of spatial dimensionality on the asymptotic
wealth distribution in additive exchange processes.  While the rate
equations apply for perfectly mixed traders or for diffusing traders in
infinite spatial dimension, $d=\infty$, deviations from the resulting
mean-field predictions are expected when $d$ is below an upper critical
dimension $d_c$.  In arbitrary spatial dimension, we assume that an
interaction occurs whenever two diffusing individuals meet.  This
transport mechanism should not be interpreted as diffusion in real
space, but rather in a space of economic activity.  For example, if
economic activity were confined to a one-dimensional road, we would use
$d=1$.  More generally, an economic network may have an effective
dimension $d>1$, with $d\to\infty$ in the modern global economy.

We further make the simplifying assumption that the diffusivity is
independent of an individual's capital.  Then an interaction occurs when
${\cal N}\cdot N\approx 1$, where ${\cal N}(\tau)$ is the average number
of distinct sites visited by a random walk in a time interval $\tau$.
This quantity scales as \cite{feller}

\begin{equation}
\label{dif}
{\cal N}(\tau)\sim\cases{\tau^{d/2},      &$d<2$;\cr                        
                         \tau/\ln\tau,    &$d=2$;\cr  
                         \tau,            &$d>2;$\cr}
\end{equation}
as $\tau\to\infty$ and thus gives the following estimates for the
density dependence of the time interval between events

\begin{equation}
\label{col}
\tau\sim \cases{N^{-2/d},            &$d<2$;\cr
                N^{-1}\ln(1/N),      &$d=2$;\cr
                N^{-1},              &$d>2$.\cr}
\end{equation}
Since the total density decreases only in events which involve the
poorest individuals, we have

\begin{equation}
\label{tot}
{dN\over dt}\sim -{c_1\over \tau}          
\end{equation}
Since we already know how the collision time $\tau$ depends on $N$, we
need to express $c_1$ on $N$ to solve Eq.~(\ref{tot}) and complete the
solution.

For random exchange, rate equations similar to Eq.~(\ref{ckran}) should
apply, except for the obvious change of the collision rate $N$ by the
dimension-dependent rate $\tau^{-1}$ from Eq.~(\ref{col}).  Thus
introducing the modified time variable

\begin{equation}
\label{Ttau}
T=\int_0^t {dt'\over \tau(t')}
\end{equation}
reduces the governing equations to the pure diffusion equation, as in
Sec.\ II.  Combining Eqs.~(\ref{NT}), (\ref{col}), and (\ref{Ttau}), we
find

\begin{equation}
\label{Nran}
N(t)\sim \cases{t^{-d/2(d+1)},        &$d<2$;\cr
             (t/\ln t)^{-1/3},  &$d=2$;\cr
             t^{-1/3},             &$d>2$.\cr}
\end{equation}
The wealth distribution is given by $c_k(t)\sim N^2x\,e^{-x^2}$ with
$N(t)$ given by Eq.~(\ref{Nran}).  In particular, the density of the
poorest individuals is proportional to $N^3$.

For greedy exchange in $d<2$, we assume that the scaling ansatz,
Eq.~(\ref{scal}), still applies, but with the slightly stronger addition
condition ${\cal C}(0)>0$.  This immediately gives $c_1\sim N^2$.  Using
this result, together with Eq.~(\ref{col}) in Eq.~(\ref{tot}), gives

\begin{equation}
\label{N}
N\sim \cases{t^{-d/(d+2)},         &$d<2$;\cr
             (t/\ln t)^{-1/2},     &$d=2$;\cr
             t^{-1/2},             &$d>2$.\cr}
\end{equation}

Our results for $N(t)$ given in Eqs.~(\ref{Nran}) and (\ref{N}) also
indicate that $d_c=2$ is the upper critical dimension for additive
capital exchange since it demarcates dimension-independent and
dimension-dependent kinetics.

\section{SUMMARY AND DISCUSSION}

We have investigated the dynamical behavior of wealth distributions in
simple capital exchange models.  For additive processes, the amount of
capital exchanged in all transactions is fixed, with the sense of the
exchange being either random -- ``random'' exchange -- or favoring the
rich -- ``greedy'' exchange.  The former leads to a Gaussian wealth
distribution, while the latter gives rise to a Fermi-like distribution.
In both cases, the number of economically viable individuals decays as a
power law in time and their average wealth correspondingly increases.  A
``very greedy'' process was also introduced in which the trading rate
between two individuals is an increasing function of their capital
difference.  This case also gives rise to a Fermi-like wealth
distribution, but with a faster decay in the density of active traders
and a concomitant faster growth in their wealth.

We next considered multiplicative processes in which a finite fraction
of the capital of one of the traders is exchanged in a transaction,
namely, $(x,y)\to(x-\alpha x, y+\alpha x)$.  From our naive viewpoint,
this multiplicative rule, especially for relatively small $\alpha$,
appears to provide a plausible description of real economic
transactions.  This process gives rise to non-local rate equations which
we have been unable to solve in closed form.  Nevertheless, considerable
insight was gained by analysis of the moments and asymptotics of the
wealth distribution.  For random exchange, the basic interaction biases
individuals with extreme wealth or extreme poverty towards the center of
the distribution and a steady state is quickly reached.  The large
wealth tail of the distribution is exponential, while at small wealth
there is a power-law tail which may be either divergent of vanishing,
for $\alpha>1/2$ or $\alpha<1/2$, respectively.

For greedy exchange, the wealth distribution assumes a power law form
$c(x,t)\simeq 1/(xt)$ for wealth in the range $(1-\alpha)^t<x<t$.  The
two cutoffs correspond to the wealth levels of the poorest and richest
individuals, levels which continue to evolve with time.  This evolving
power law form conforms to our uninformed view of the wealth
distribution in certain countries with developing economies.  If we
define the rich as those whose capital is greater than the average
initial capital, their fraction decays as $\ln t/t$ and their average
wealth grows as $t/\ln t$

Formally, it would be interesting to investigate limiting situations
where well-known pathologies in the underlying capital exchange process
can occur.  For example, as briefly mentioned in Sec.\ II, there is the
possibility of one individual acquiring a finite fraction of the total
capital in a finite time.  This feature corresponds to gelation in
polymerization processes\cite{gel}.  Mathematically, this singularity
should manifest itself in the violation of wealth conservation, where
the loss of wealth in the population of finitely wealthy individuals
signals the appearance of an infinitely rich individual.  For
multiplicative exchange processes which possess two conservation laws --
total density and wealth -- there is also the possibility of losing
population of individuals of finite wealth to a ``dust'' phase which
consists of a finite fraction of all individuals who possess no wealth.
This latter phenomenon is analogous to the ``shattering'' transition in
fragmentation processes\cite{fil}.

At a practical level, a potentially fruitful direction would be to
incorporate additional realistic elements into capital exchange models.
For example, in all societies, there is some form of wealth
redistribution by taxation.  It would be worthwhile to determine how
various types of taxation algorithms -- graduated tax, ``flat'' tax,
{\it etc.}, in concert with capital exchange -- influence the asymptotic
wealth distribution.  Conversely, it would be interesting to understand
how financial ``safety nets'', such as welfare, affect the density and
wealth of the poorest individuals.  Another aspect worth pursuing is
heterogeneous exchange rules, where the factor $\alpha$ is either
different for each individual or depends on some other aspect of a
trading event.  One might hope to find universality in the wealth
distributions with respect to this heterogeneity.

More generally, the focus on conserved exchange of capital is deficient
because there is no mechanism for economic growth.  Pure exchange might
be appropriate for short time scales during which economic growth is
negligible.  However, pure exchange cannot be expected to be suitable on
longer time scales where long-term economic growth is an overriding
influence.  Thus a two-body interaction rule with a net increase in the
capital might be a more appropriate long-time description of an
economically interacting system.
 
Perhaps the simplest realization is random capital growth, $(j,k)\to
(j+1,k)$ or $(j,k+1)$.  Within the rate equation approach, this model is
readily solved to find the asymptotic wealth distribution $c_k(t)\sim
t^{-1/2} \exp[-(t-k)^2/2t]$.  Hence the average wealth grows linearly in
time, $\langle k\rangle\simeq t$, while the relative fluctuation
${\sqrt{\langle (\Delta k)^2\rangle}/\langle k\rangle}$ decreases as
$t^{-1/2}$.  Thus an (economically) fair society with a sharp wealth
distribution arises, albeit with an unrealistic linear, rather than
exponential, growth of the average wealth.  Such an exponential growth
can be achieved, {\it e.\ g.}, by an interaction rate which equals the
sum of capitals of the participants, $K(i,j)=i+j$.  the total wealth
density $M=\sum kc_k$ now obeys $\dot M=2NM$ and thus grows
exponentially.  This model fulfills the Marxist dream of fast wealth
growth $\langle k\rangle\sim e^{2Nt}$ with all participating equally in
the prosperity (the relative fluctuation decreasing as
$\sqrt{t}\,e^{-Nt}$).  For an interaction rate which increases even more
rapidly with wealth, there is a pathology analogous to gelation --
infinite prosperity in a finite time.  For example, for the product
kernel, $K(i,j)=ij$, the wealth distribution for a monodisperse initial
condition is $c_k(t)=(1-t)t^{k-1}$.  Interestingly, the Marxist ideal
lies on the boundary between algebraic wealth growth and the pathology
of a finite-time wealth divergence, a feature similar to the ``life on
the edge of chaos'' advocated by Kauffman as a generic property of
complex systems\cite{kauf}.

Related behavior occurs in a multiplicative growth process where
$(x,y)\to (x,y+\alpha x)$.  Naively, a single multiplicative interaction
which changes the capital by an amount proportional to the existing
capital $k$ is roughly equivalent to $k$ successive additive
interactions.  Consequently, this simple version of multiplicative
growth should exhibit exponential wealth growth.  For the specific case
that we were able to solve, $\alpha=1$, the wealth distribution
approaches the scaling form $c_k(t)\sim e^{-t}\exp(-x e^{-t})$.
However, the wealth distribution is broad, a feature generic of
multiplicative growth processes for general values of $\alpha$.  

More realistically, an overall growth process should incorporate the
possibility that in a single trade: both traders profit, one profits, or
neither profits, but with an overall gain in capital when averaged over
many trading events.  It would be interesting to determine the nature of
the wealth distribution in such a multiplicative system.

\section{ACKNOWLEDGMENTS}

We thank J. L. Spouge for stimulating discussions which helped lead to
the formulation of the models discussed in this work.  We also thank
D.~Ray and R.~Rosenthal of the Boston University Economics Department for
helpful advice.  This research was supported in part by the NSF (grant
DMR-9632059), and by the ARO (grant DAAH04-96-1-0114).  This financial
assistance is gratefully acknowledged.

{\em Added Note:} After this work was completed, we learned on a
related work\cite{tak} in which a model similar to greedy
multiplicative exchange was employed to describe the distribution of
company sizes.  We thank H. Takayasu for kindly informing us of this
development.

\end{multicols} 
\end{document}